\documentclass[12pt]{iopart}
\usepackage{dcolumn}
\usepackage{bm}
\usepackage[sort&compress]{natbib}
\bibpunct{[}{]}{,}{a}{}{;}
\def\newblock{\hskip .11em plus .33em minus .07em}
\usepackage{graphicx}
\usepackage{epsf}

\usepackage{amssymb,textcomp}
\usepackage{threeparttable,booktabs}

\usepackage{caption}

\usepackage{array}
\newcommand{\PreserveBackslash}[1]{\let\temp=\\#1\let\\=\temp}
\newcolumntype{C}[1]{>{\PreserveBackslash\centering}p{#1}}
\newcolumntype{R}[1]{>{\PreserveBackslash\raggedleft}p{#1}}
\newcolumntype{L}[1]{>{\PreserveBackslash\raggedright}p{#1}}

\makeatletter
\newcommand{\rmnum}[1]{\romannumeral #1}
\newcommand{\Rmnum}[1]{\expandafter\@slowromancap\romannumeral #1@}
\makeatother

\usepackage{ifpdf}
\ifpdf
\usepackage[pdftex]{hyperref}
\else
  \usepackage[dvipdfm]{hyperref}
\fi
 \hypersetup{bookmarksnumbered,%
            colorlinks,%
               linkcolor=blue,%
               citecolor=blue,%
              plainpages=false,%
            pdfstartview=FitH}

\begin{document}
\pagestyle{plain}
\title[]{First-principles study of the structural, elastic, and electronic properties of the cubic perovskite BaHfO$_3$}

\author{Hongsheng Zhao$^{1,2,3}$, Aimin Chang$^1$\footnote{Corresponding author.} and Yunlan Wang$^{4}$}

\address{$^1$ Xinjiang Technical Institute of Physics and
Chemistry, Chinese Academy of Sciences,
Urumqi 830011, P.R.China}
\address{$^2$ Institute of Semiconductors, Chinese Academy of
Sciences
P.O.Box 912, Beijing 100083, P.R.China}
\address{$^3$ Graduate School of Chinese Academy of Sciences,
Beijing 100049, P.R.China}
\address{$^4$ Center for High Performance Computing, Northwestern Polytechnical University, Xi'an 710072, P.R.China}
\ead{\href{mailto:zhaohscas@yahoo.com.cn}{zhaohscas@yahoo.com.cn}}

\begin{abstract}
First principles study of structural, elastic, and electronic properties of the cubic perovskitetype
BaHfO$_3$ has been performed using the plane wave ultrasoft pseudo-potential method based on density functional theory  with revised Perdew-Burke-Ernzerhof exchange-correlation functional of the generalized gradient approximation (GGA-RPBE). The calculated equilibrium lattice constant of this compound is in good
agreement with the available experimental and theoretical data reported in the
literatures. The
independent elastic constants (\emph{C}$_{11}$, \emph{C}$_{12}$, and \emph{C}$_{44}$), bulk modules \emph{B} and its pressure derivatives $B^{\prime}$, compressibility $\beta$,
shear modulus \emph{G}, Young's modulus \emph{Y}, Poisson's ratio $\nu$, and Lam\'{e} constants ($\mu, \lambda$) are obtained and analyzed in comparison
with the available theoretical and experimental data for both the singlecrystalline and polycrystalline BaHfO$_3$. The band structure calculations show that BaHfO$_3$ is
a indirect bandgap material (R-$\Gamma$ = 3.11 eV) derived basically from the occupied O 2\emph{p} and unoccupied Hf 5\emph{d} states, and it still awaits experimental confirmation. The density of states (total,
site-projected, and \emph{l}-decomposed) and the bonding charge
density calculations make it clear that the covalent bonds exist between the Hf and O atoms and the ionic bonds exist between the Ba atoms and HfO$_3$ ionic groups in BaHfO$_3$. From our calculations, it is shown that BaHfO$_3$ should be promising as a candidate for synthesis
and design of superhard materials due to the covalent bonding between the transition metal Hf 5\emph{d} and O 2\emph{p} states.

\end{abstract}
\submitto{\JPCM}
\maketitle
\normalsize

\section{Introduction}
\label{intro}

The perovskite-type oxides have the potential to be attractive
functional materials because they have a wide range of interesting physical properties which have been extensively reported, such as ferroelectricity \citep{Samantaray2004,Bednorz1984,Samantaray2005}, semiconductivity
\cite{Frederikse1964}, superconductivity \cite{Koonce1967}, catalytic activity \cite{Henrich1985}, piezoelectricity \cite{Wang2007,Baettig2005}, colossal magnetoresistance \cite{Millis1996,Tokura2000}, and
thermoelectricity \cite{Muta2003,Henrich1985}.
Up to now, the perovskite-type oxides have been widely used in many fields,
including spintronic devices, optical
wave guides, laser-host crystals, high temperature oxygen sensors,
surface acoustic wave devices, non-volatile memories, dynamic
random access memories, frequency doublers, piezoelectric actuator
materials, catalyst electrodes in certain types of fuel cells, and high-$\kappa{}$ capacitors in various applications \cite{Mete2003,Henrich1994}.

\begin{figure}[htbp]
  \centering
  \includegraphics[height=5cm]{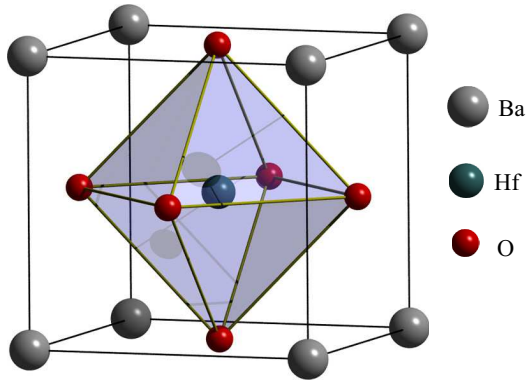}
  \captionsetup{margin={1cm,0cm}}
  \caption{(Color online) The crystal structural perspective polyhedral view of the cubic perovskite-type BaHfO$_3$.}
  \label{structure_BaHfO_3}
\end{figure}

The perovskite-type alkaline earth hafnate, BaHfO$_3$
 belongs to the family of the perovskite-type
oxides ABO$_3$ and is assigned to the pseudocubic
cell of the perovskite structure, where A and B cations are in
12-fold and 6-fold coordination, respectively.
More specifically, a Ba atom sits at cube corner
position (0, 0, 0), a Hf atom sits at
body centre
position (1/2, 1/2, 1/2), and three O atoms sit at face
centred positions (1/2, 1/2, 0) forming a regular octahedron, as depicted in \Fref{structure_BaHfO_3}.
There are few reported experimental studies devoted to this compound.
Very recently, a polycrystalline
sample of BaHfO$_3$ has been
successfully prepared and some physical properties of this compound
have been investigated\cite{Maekawa2006}.

A more complete understanding of the physical properties of BaHfO$_3$ is prerequisite to
the eventual technological applications of this compound. A systematic theoretical investigations
of the structural, elastic and electronic properties of BaHfO$_3$ is
necessary. As far as we know, there is surprisingly little theoretical work exploring its physical properties \cite{Kitamura1998,Bouhemadou2008,Xiao2007}.

In this paper, we
report a systematic study of the structural, elastic, and electronic properties of the cubic perovskite BaHfO$_3$, using the state-of-the-art pseudo-potential plane-waves (PP-PW) approach based on density functional theory (DFT) with revised Perdew-Burke-Ernzerhof functional of the generalized gradient approximation
(GGA-RPBE). Our studies have
been motivated by the following reasons.

Even though the mentioned experimental and theoretical results \cite{Bouhemadou2008,Xiao2007,Maekawa2006,Kitamura1998} for
BaHfO$_3$ have been published, some results are still controversial and remain open problems.
In the calculation of bulk modulus by Bouhemadou \etal \cite{Bouhemadou2008}, a uniform \emph{P}-\emph{V} data set was used to fit to the third order
Birch-Murnaghan equation \cite{Poirier1991}, which would give inaccurate results.
The calculated bandgap results of BaHfO$_3$ by Xiao \etal \cite{Xiao2007} and Bouhemadou et al \cite{Bouhemadou2008} are incompatible.
Theoretical investigations of the bonding charge densities properties of BaHfO$_3$ hasn't been done. Therefore, based on DFT-GGA-RPBE method, we calculated the bulk modulus by using a dense sampling technology in the low-pressure region, the band structures,
density of states (total, site-projected,
and l-decomposed) and the bonding charge densities. It allows us to understand and clarify the uncertainties in comparison with other experimental and theoretical results \cite{Bouhemadou2008,Xiao2007,Maekawa2006,Kitamura1998}.

Elastic behavior for cubic perovskite BaHfO$_3$ is
of great interest for its potential applications. Therefore,
we have predicted the elastic parameters from accurate first principles
DFT-GGA-RPBE calculations not only for
BaHfO$_3$ monocrystal but also for its polycrystalline
state, as this material has been prepared as polycrystalline samples \cite{Zhang1993,Maekawa2006}.
In this study, a set of physical parameters of monocrystalline
BaHfO$_3$, such as optimized lattice parameter, elastic
constants \emph{C}$_{11}$, \emph{C}$_{12}$, and \emph{C}$_{44}$, bulk modulus \emph{B}, compressibility
$\beta$, and shear modulus \emph{G}, is
calculated, and the numerical estimates of the bulk modulus \emph{B}, compressibility
$\beta$, and shear modulus \emph{G}, Young's modules \emph{Y}, Poisson's ratio ($\nu$), and Lam\'{e} constants ($\mu, \lambda$) of the polycrystalline BaHfO$_3$ (in the framework of the Voigt-Reuss-Hill approximation) are obtained and analyzed.

This paper is organized as follows: In Section \ref{compdetails} we briefly described
the computational techniques used in the present work. Results
and discussions of the structural, elastic, and
electronic properties will be presented in Section \ref{res_dis}. Finally, conclusions and
remarks are given in Section \ref{conclu}.

\section{Computational details}
\label{compdetails}

In this paper, all calculations
are performed with the Cambridge Serial Total Energy Package (CASTEP)
code \cite{Segall2002}, which is a implementation in pseudo-potential plane-waves (PP-PW) based density functional theory (DFT). The exchange-correlation potential
is treated with the GGA-RPBE functional, developed by Hammer \etal \cite{Hammer1999}.
Coulomb potential energy
caused by electron-ion interaction is described using Vanderbilt-type ultrasoft
scheme \cite{Laasonen1991,Laasonen1993,Vanderbilt1990}. The Ba
5\emph{s}$^2$, Ba 5\emph{p}$^6$, Ba 6\emph{s}$^2$, Hf 5\emph{d}$^2$, Hf 6\emph{s}$^2$, O 2\emph{s}$^2$,
and O 2\emph{p}$^6$ electrons are treated as valence electrons for performing the pseudo atomic calculation. A 380~eV energy cut-off for the plane-wave basis and a
(8, 8, 8) mesh in the Monkhorst-Pack scheme for the Brillouin-zone sampling are adopted during structure optimization. Atomic positions of BaHfO$_3$ are relaxed and optimized with a
density mixing scheme using the Broyden-Fletcher-Goldfarb-Shanno (BFGS) minimization technique. The following convergence criterions are reached: energy
change per atom less than $2 \times 10^{-6}$~eV, residual force less
than 0.01~eV/{\AA}, stress below 0.05 GPa, and the displacement of
atoms during the geometry optimization less than $5 \times 10^{-4}${\AA}.

With the equilibrium geometry configuration,
we applied the so-called stress-strain method to
obtain the elastic constants in which the stress can be easily
obtained within the density functional based electronic structure
method \cite{Nielsen1985}. The stress-strain relation can be described as \cite{Fan2006}

\begin{equation}
(\sigma _1, \sigma _2, \sigma _3, \sigma _4, \sigma _5, \sigma _6) = C(\varepsilon _1, \varepsilon _2, \varepsilon _3, \varepsilon _4, \varepsilon _5, \varepsilon _6)^T ,
\label{stress-strain}
\end{equation}

\noindent where \emph{C} is the elastic stiffness matrix.
For the cubic crystal, there are only three non-zero independent
symmetry elements (\emph{C}$_{11}$, \emph{C}$_{12}$, and \emph{C}$_{44}$), each of which represents three equal elastic constants (\emph{C}$_{11}$=\emph{C}$_{22}$=\emph{C}$_{33}$; \emph{C}$_{12}$=\emph{C}$_{23}$=\emph{C}$_{31}$; \emph{C}$_{44}$=\emph{C}$_{55}$=\emph{C}$_{66}$) \cite{Nye1985}. A single strain with non-zero first and fourth components can give stresses relating to all three of these coefficients, yielding a very efficient method for obtaining elastic constants for the cubic system.

\section{Results and discussions}
\label{res_dis}

\subsection{Lattice constants}
\label{ssec:Lattice-constan}

First, the equilibrium lattice constants (\emph{a}$_0$) for the ideal
stoichiometric perovskite BaHfO$_3$ was calculated.
The results are listed in Table \ref{tab:elastic_BaHfO3}, along with the available experimental and theoretical data. As can be seen, our calculated equilibrium lattice
parameter (\emph{a}$_0$) is in excellent agreement with the experimental
data and previous calculations: the calculated lattice constant deviates from the measured
and the calculated ones within 2.8\% and 0.8\% respectively. The above results also show that the computational methods and parameters used in this paper are reasonable.

\subsection{Elastic properties}
\label{ssec:Elastic-propert}

Within the framework of the DFT-GGA-RPBE calculations,
the values of elastic constants
\emph{C}$_{ij}$ for BaHfO$_3$ are presented in Table \ref{tab:elastic_BaHfO3}.
These three independent elastic constants in a cubic symmetry
(\emph{C}$_{11}$, \emph{C}$_{12}$, and \emph{C}$_{44}$) were estimated by calculating the
stress tensors on applying strains to an equilibrium structure.
All \emph{C}$_{ij}$ constants for BaHfO$_3$ crystal are positive and
satisfy the generalized criteria \cite{Yip2001,Grimvall1986,Wang1993} for mechanically stable
crystals: (\emph{C}$_{11}$-\emph{C}$_{12}$)$>$0, (\emph{C}$_{11}$+2\emph{C}$_{12}$)$>$0, and \emph{C}$_{44}>$0.

\begin{table}[h]
\captionsetup{margin={1.5cm,0cm}}
\caption{The lattice parameter (\emph{a}$_0$, in {\AA}), elastic constants
(\emph{C}$_{ij}$, in GPa), bulk modulus (\emph{B}, in GPa) and its pressure derivatives ($B^{\prime}$), compressibility ($\beta$, in
GPa$^{-1}$),
shear modulus (\emph{G}, in GPa), Young's modulus (\emph{Y}, in GPa), Poisson's ratio ($\nu$), and Lam\'{e} constants ($\mu, \lambda$) for BaHfO$_3$. }
\label{tab:elastic_BaHfO3}
\begin{center}
\begin{threeparttable}
\lineup\begin{tabular}{@{\extracolsep{\fill}}llll@{\extracolsep{\fill}}}
\toprule
Parameters                                                               & Present work                       & Experimental                       & Other calculations       \\
\midrule
\emph{a}$_0$    & \0\04.2858                         & \0\04.171\tnote{a}                 & \hspace{1cm}\0\04.2499\tnote{b} \\
                &                                    & \0\04.17\tnote{c}                  & \hspace{1cm}\0\04.310\tnote{d}  \\
\emph{C}$_{11}$ & 313.18                             & ---                                & \hspace{1cm}422\tnote{b}        \\
                &                                    &                                    & \hspace{1cm}317.5\tnote{d}      \\
\emph{C}$_{12}$ & \063.49                            & ---                                & \hspace{1cm}\080\tnote{b}       \\
                &                                    &                                    & \hspace{1cm}\026.6\tnote{d}     \\
\emph{C}$_{44}$ & \070.33                            & ---                                & \hspace{1cm}\073\tnote{b}       \\
                &                                    &                                    & \hspace{1cm}\074.3\tnote{d}     \\
\emph{B}        & 146.72\tnote{e}                    & ---                                & \hspace{1cm}194\tnote{b,e}      \\
                & 145.50\tnote{f}                    &                                    & \hspace{1cm}186\tnote{b,f}      \\
                &                                    &                                    & \hspace{1cm}123.6\tnote{d,e}    \\
\emph{B}$^{\prime}$ &\0\05.08\tnote{f} & --- & \hspace{1cm}\0\05.02\tnote{b,f}\\
$\beta$         & \0\06.82$ \times 10^{-3}$\tnote{e} & \0\08.78$ \times 10^{-3}$\tnote{a} & \hspace{1cm}---                 \\
                & \0\06.87$ \times 10^{-3}$\tnote{f} &                                    &                                 \\
\emph{G}        & \088.68\tnote{e}                   & \079.5\tnote{a}                    & \hspace{1cm}\097.6\tnote{d}     \\
\emph{Y}        & 221.42\tnote{e}                    & 194\tnote{a}                       & \hspace{1cm}---                 \\
$\nu$           & \0\00.2485                         & ---                                & \hspace{1cm}---                 \\
$\mu$           & \088.68                            & ---                                & \hspace{1cm}---                 \\
$\lambda$       & \087.60                            & ---                                & \hspace{1cm}---                 \\
\bottomrule
\end{tabular}
\begin{tablenotes}
\footnotesize
\item[a] Reference \cite{Maekawa2006}.
\item[b] Reference \cite{Bouhemadou2008}.
\item[c] Reference \cite{L'opez1991}.
\item[d] Reference \cite{Xiao2007}.
\item[e] Obtained from the Voigt-Reuss-Hill (VRH) approximation \cite{Grimvall1986}.
\item[f] Obtained from fitting the third-order Birch-Murnaghan equation \cite{Poirier1991}.
\end{tablenotes}
\end{threeparttable}
\end{center}
\end{table}

Considering that the above elastic parameters are obtained from
first-principles calculations of BaHfO$_3$
monocrystal. However, this material has been prepared
as polycrystalline structure, \cite{Maekawa2006,Zhang1993} and, therefore, it is useful to
evaluate the corresponding parameters for the polycrystalline
samples. For this purpose we have utilized the Voigt-Reuss-Hill (VRH) approximation
\cite{Grimvall1986}. In this approach, the actual
effective modulus for polycrystals could be approximated by
the arithmetic mean of the two well-known bounds for
monocrystals according to Voigt \cite{Voigt1928} and Reuss \cite{Reuss1929}. Then, the
main mechanical parameters for cubic perovskite BaHfO$_3$,
namely, bulk modulus (\emph{B}), compressibility ($\beta$), shear modulus (\emph{G}), Young's modulus
(\emph{Y}), Poisson's ratio ($\nu$), and Lam\'{e} constants ($\mu, \lambda$) in the
two mentioned approximations, Voigt (V) \cite{Voigt1928} and Reuss (R) \cite{Reuss1929},
are calculated from the elastic constants of single crystal BaHfO$_3$ according to
the following formulae \cite{Anderson1963,Schreiber1973}:
\numparts
\begin{eqnarray}
B_{{\rm{V,R,VRH}}} & =\left({C_{11}+2C_{12}}\right)/3. \label{eqn:b_vrh}                                                                                               \\
G_{\rm{V}}         & =(C_{11}-C_{12}+3C_{44})/5, \cr
G_{\rm{R}}         & =5/\left[{4\left({S_{11}-S_{12}}\right)+3S_{44}}\right] \cr
                   & ={\rm{5}}\left({C_{11}-C_{12}}\right)C_{44}{\rm{/}}\left[{{\rm{4}}C_{44}{\rm{ + 3}}\left({C_{11}-C_{12}}\right)}\right], \cr
G_{{\rm{VRH}}}     & = \left( {G_{\rm{V}}  + G_{\rm{R}} } \right)/2. \label{eqn:g_vrh}                                                                                 \\
Y_{{\rm{VRH}}}     & = 9G_{{\rm{VRH}}} B_{{\rm{VRH}}} /\left( {G_{{\rm{VRH}}}  + 3B_{{\rm{VRH}}} } \right). \label{eqn:y_vrh}                                          \\
\nu                & = \left( {3B_{{\rm{VRH}}}  - 2G_{{\rm{VRH}}} } \right)/\left[ {2\left( {3B_{{\rm{VRH}}}  + G_{{\rm{VRH}}} } \right)} \right]. \label{eqn:Poisson} \\
\mu                & = Y_{{\rm{VRH}}} {\rm{/}}\left[ {{\rm{2}}\left( {{\rm{1 + }}\nu } \right)} \right], \cr
\lambda            & = \nu Y_{{\rm{VRH}}} /\left[ {\left( {{\rm{1 + }}\nu } \right)\left( {{\rm{1 - 2}}\nu } \right)} \right]. \label{eqn:lame}
\end{eqnarray}
\endnumparts

\noindent Where \emph{S} in \eref{eqn:g_vrh} is the elastic compliance matrix, which is equal to the inverse of the elastic stiffness matrix \emph{C} in \eref{stress-strain}.

The calculated values of the elastic parameters
for the polycrystalline BaHfO$_3$ are given in \Tref{tab:elastic_BaHfO3}. Our calculated elastic parameters are in reasonably good agreement with most of the available experimental data. The calculated large Bulk Modulus indicates a relative high incompressibility and hardness of BaHfO$_3$ according to the strong correlation between the bulk
modulus and hardness of materials \cite{Haines2001}, implying that BaHfO$_3$ should be promising as a candidate for synthesis and design of superhard materials.

\begin{figure}[h]
  \centering
  \includegraphics[height=5cm]{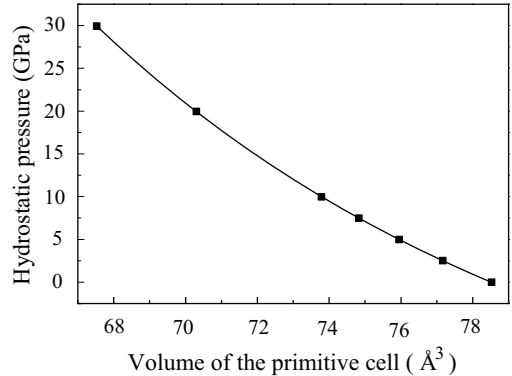}
  \captionsetup{margin={1cm,0cm}}
  \caption{ The calculated pressure-volume relation for BaHfO$_3$. The solid line is given
by the third-order Birch-Murnaghan equation of state.}
\label{fig:eos_pv}
\end{figure}

In order to further validate the reliability and accuracy of
our calculated elastic constants for BaHfO$_3$, the calculated unit cell volumes under a series of applied hydrostatic
pressures were used to
construct the \emph{P}-\emph{V} data set, which was subsequently fitted to a third-order Birch-Murnaghan equation \cite{Poirier1991}
 (\Fref{fig:eos_pv}):

\begin{equation}
P=\displaystyle
\frac{{3B}}{2}\left[{\left({\frac{{V_{0}}}{V}}\right)^{7/3}-
\left({\frac{{V_{0}}}{V}}\right)^{5/3}}\right]\left\{{1+\frac{3}{4}
\left({B^{\prime}-4}\right)\left[{\left({\frac{{V_{0}}}{V}}\right)^{2/3}-1}\right]}\right\},
\end{equation}

\noindent where \emph{V}$_0$ is the the primitive unit cell's volume of BaHfO$_3$ determined from the zero pressure, \emph{B}$_0$ is the bulk modulus and \emph{B}$^{\prime}$
is its pressure derivative at zero pressure.

Considering that the values of \emph{B} and \emph{B}$^{\prime}$ determined by the equation of state (EOS) fitting process depend on the pressure range used in the calculations. Experimental values obtained using a diamond anvil cell are usually in the range 0$\sim$30 GPa, so this pressure range is adopted by the present work. In addition, more accurate sampling in the low-pressure region is used to obtain an accurate value of the bulk modulus.

We obtained, by least-squares fitting of the non-uniform \emph{P}-\emph{V} data set to the third-order Birch-Murnaghan equation \cite{Poirier1991}, the bulk modulus \emph{B}$_0$ and
its pressure derivative \emph{B}$^{\prime}$ at zero pressure. The calculated results are listed in \Tref{tab:elastic_BaHfO3}. Comparing the value from the Voigt-Reuss-Hill (VRH) approximation \cite{Grimvall1986}, our result is more accurate than that by Bouhemadou \etal \cite{Bouhemadou2008}.

\subsection{Electronic properties}
\label{ssec:Electro-propert}

\begin{figure}[h]
  \centering
  \includegraphics[height=7.2cm]{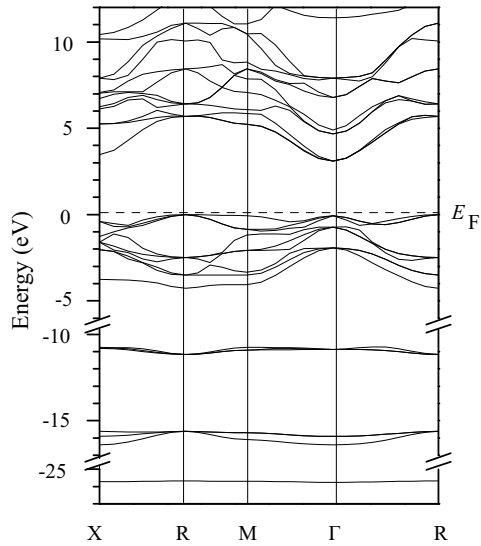}
  \captionsetup{margin={1cm,0cm}}
  \caption{Band structure for BaHfO$_3$ along the
high symmetry lines of the Brillouin
zone, Fermi level $E_{\rm{F}}$=0~eV.}
    \label{fig:bandstru}
\end{figure}

\begin{figure}[h]
  \centering
  \includegraphics[height=10cm]{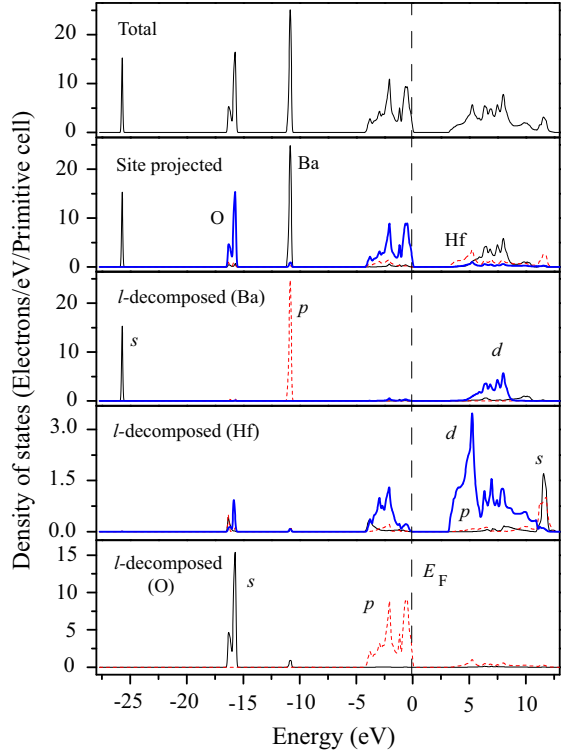}
  \captionsetup{margin={1cm,0cm}}
 \caption{(Color online) Total, site-projected, and \emph{l}-decomposed DOS for BaHfO$_3$. The black (thin), red (dashed), and blue (thick) lines represent the Ba, Hf, and O atoms' DOS contributions in the site-projected case and  the \emph{s}, \emph{p}, and \emph{d} orbitals' DOS contributions in the \emph{l}-decomposed case, respectively, Fermi level $E_{\rm{F}}$=0~eV.}
    \label{fig:dos}
   \end{figure}

\begin{figure}[ht]
  \centering
  \includegraphics[height=5cm]{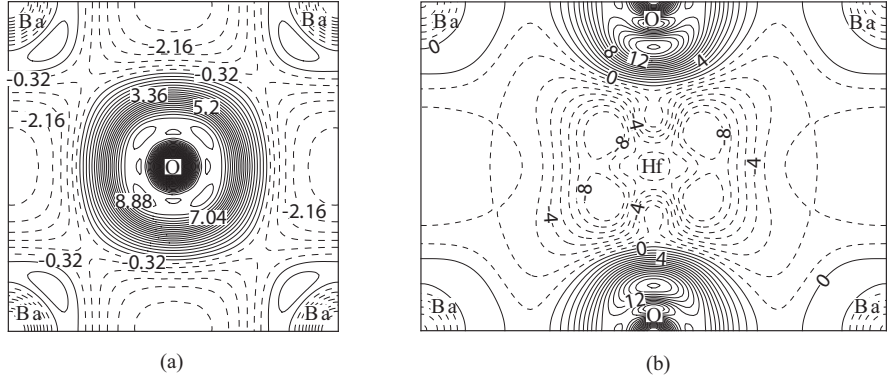}
  \captionsetup{margin={1cm,0cm}}
  \caption{The bonding
charge densities for BaHfO$_3$: (a) in the (100) plane and (b) in the (110) plane.
The contour step size is 0.46 e/f.u. and
1 e/f.u. for (a) and (b), respectively, where f.u. denotes formula unit.}
  \label{fig:charge_density}
  \end{figure}

The electronic band structure, density of states (total, site-projected, and \emph{l}-decomposed) and bonding charge
density which have
been calculated for the equilibrium geometry of
BaHfO$_3$ are shown in \Fref{fig:bandstru}, \Fref{fig:dos} and \Fref{fig:charge_density}, respectively.

The calculated energy band structure of BaHfO$_3$ along the
high symmetry lines in the first Brillouin zone is shown in \Fref{fig:bandstru}.
The Fermi level $E_{\rm{F}}$ is chosen to locate at 0~eV and coincides with the top of the valence
band. The valence band maximum (VBM) and conduction band minimum (CBM) are found to be 0~eV at R point and 3.11~eV at $\Gamma$ point, respectively, indicating that this compound is a indirect bandgap material (R-$\Gamma$). These results are in good agreement with those obtained by Xiao \etal \cite{Xiao2007}, but inconsistent with those by Bouhemadou \etal \cite{Bouhemadou2008}. Considering that no experimental bandgap data for BaHfO$_3$ are available, the experimental studies are necessary to further testify the theoretical results. The values of the main band gaps and the width of the valence band obtained in the present work along with other theoretical results of the band
structure for this compound are given in \Tref{tab:bandstru_data}.

\begin{table}[h]
\captionsetup{margin={1.5cm,0cm}}
\caption{Electron band eigenvalues for the lowest conduction bands and the upper
valence band widths of the BaHfO$_3$ (all of the
energies are in eV).}
\label{tab:bandstru_data}
\begin{center}
\begin{threeparttable}
\lineup\begin{tabular}{@{\extracolsep{\fill}}lll@{\extracolsep{\fill}}}
\toprule
Band structure features  & This work & Other calculations \\
\midrule
          Direct         &           &                                \\
  $\Gamma$-$\Gamma$    & \03.17    & \hspace{1cm}2.99\tnote{a}      \\
          M-M          & \05.30    & \hspace{1cm}5.61\tnote{a}      \\
          R-R          & \05.76    & \hspace{1cm}5.87\tnote{a}      \\
          X-X          & \03.87    & \hspace{1cm}3.80\tnote{a}      \\
           Indirect      &           &                                \\
          $\Gamma$-M   & \05.30    & \hspace{1cm}5.50\tnote{a}      \\
          $\Gamma$-R   & \05.82    & \hspace{1cm}5.83\tnote{a}      \\
          $\Gamma$-X   & \03.54    & \hspace{1cm}3.42\tnote{a}      \\
           M-$\Gamma$  & \03.17    & \hspace{1cm}---                \\
           R-$\Gamma$  & \03.11    & \hspace{1cm}3.1\tnote{b}       \\
           X-$\Gamma$  & \03.49    & \hspace{1cm}---                \\
 Total-valence bandwidth & 25.95     & \hspace{1cm}---                \\
\bottomrule
\end{tabular}
\begin{tablenotes}
\footnotesize
\item[a] Reference \cite{Bouhemadou2008}.
\item[b] Reference \cite{Xiao2007}.
\end{tablenotes}
\end{threeparttable}
\end{center}
\end{table}

The total, site-projected, and \emph{l}-decomposed densities of states (DOS) for BaHfO$_3$ are presented
in \Fref{fig:dos}. The results are consistent with other theoretical studies \cite{Xiao2007,Bouhemadou2008}. The Fermi
level ($E_{\rm{F}}$) and the conduction band minimum (CBM) are located close to peaks provided mainly by O 2\emph{p} electrons and Hf 5\emph{d} electrons, respectively. The lowest region of the valence band situated in the range from $-$25.95 to $-$25.46 eV is due to the Ba 6\emph{s} states. The valence band which extend from
$-$16.59 up to $-$15.36 eV is derived basically from the O 2\emph{s} states with some mixing of Hf 6\emph{s}, Hf 5\emph{d},
and Ba 5\emph{p} states. The valence band situated in the region from $-$11.39 up to
$-$10.51 eV is essentially dominated by Ba 5\emph{p} states, with admixture
from O 2\emph{s} states. The upper valence band (between $-$4.47 and 0 eV)
is mainly due to O 2\emph{p} states hybridized with some Hf 5\emph{d} electrons,
which suggests covalent bonding contributions in BaHfO$_3$. Here Hf 5\emph{d} and O 2\emph{p} orbitals
overlap, resulting in covalent bonding between the Hf and O atoms in this material. The Hf-O
covalency will also responsible for the hardness of this compound \cite{Jhi1999}. The lower conduction bands are contributed
mainly by Hf 5\emph{d}, Ba 5\emph{d}, and Ba 6\emph{s} states.

The bonding charge
density is an useful tool to describe the redistribution
of the electrons along with the bonding process. In order to interpret the bonding mechanism of
BaHfO$_3$, the
bonding-charge density was used to study the
bonding characteristics of this compound. The bonding charge density is
defined as the difference between the total charge density in the
solid and the superposition of neutral atomic charge densities placed at the same atomic sites,
with the formula as below \cite{Sun1995}:

\begin{equation}
\Delta\rho(r)=\rho_{solid}(r)-\mathop\Sigma\limits_\alpha\rho_{\alpha}(r-r_{\alpha}),
\end{equation}

\noindent where $\rho_{solid(r)}$ is the total charge density of the equilibrium
structure and $\rho _\alpha  (r - r_\alpha  )$
 is the charge density of the neutral
atom $\alpha$. Therefore, the bonding charge density represents the
net charge redistribution as atoms are brought together to form
the crystal.

The bonding-charge densities in the (100) and (110)
plane of BaHfO$_3$ have been calculated and shown in \Fref{fig:charge_density}(a) and (b), with solid and dashed contour lines denoting positive (hence accumulation) and negative (hence depletion) of
electron density, respectively, relative to the atomic electron density. The positive (negative) charge redistribution
can be identified with electronic transfer into
(outward) bonding or anti-bonding electronic states.

As shown in \Fref{fig:charge_density}, Ba atoms have
more anti-bonding charge density in the (100)
plane than that in the (110) plane. Moreover in
the (110) plane, the electrons
accumulate and distribute along the Hf-O bond and deflect to bonding line, resulting in
strong covalent characteristics of the Hf-O bond, which coincides well with the results obtained from the densities of states (DOS) analysis.
The covalent characteristics of the Hf-O bond will also result in a high
hardness value and can contribute to the elastic
incompressibility (bulk modulus) of this material, which are consistent with the corresponding results obtained from the elastic properties and density of states (DOS) calculations, as well as the experimental data \cite{Maekawa2006}.  The electrons depletion process occurs
near the Ba atoms both in the (100) and (110) planes could be
ascribed to the presence of the strong ionic characteristics between the Ba atoms and HfO$_3$ ionic groups.

\section{Conclusions}
\label{conclu}

The structural, elastic, and electronic properties of the cubic perovskite BaHfO$_3$ have been studied by using the density functional theory with the GGA-RPBE approximation. Our main results and conclusions can be summarized as follows:

(\rmnum{1}) The calculated equilibrium lattice
constant of this compound is in good agreement
with the available experimental and theoretical
data reported in the literatures.

(\rmnum{2}) We have used a dense sampling technology in the low-pressure region to obtain an
accurate value of the bulk modulus by fitting the third-order Birch-Murnaghan equation \cite{Poirier1991}. Comparing the value from the Voigt-Reuss-Hill (VRH) approximation \cite{Grimvall1986}, our result is more accurate than that by Bouhemadou \etal \cite{Bouhemadou2008}.

(\rmnum{3}) The electronic structure calculations showed that BaHfO$_3$ is
a indirect bandgap material (R-$\Gamma$ = 3.11 eV). This result is consistent with the result by Xiao \etal \cite{Xiao2007}, but differs from that by Bouhemadou \etal \cite{Bouhemadou2008}. Considering that no experimental bandgap data for BaHfO$_3$ are available, the experimental studies are necessary to further testify the theoretical results.

(\rmnum{4}) The calculated elastic properties, including bulk modulus (\emph{B}) and its pressure derivatives ($B^{\prime}$), compressibility ($\beta$), shear modulus (\emph{G}), Young's modulus
(\emph{Y}),  Poisson's ratio ($\nu$), and Lam\'{e} constants ($\mu, \lambda$), are in reasonably good agreement with the available experimental data.

(\rmnum{5}) Analyses of the density of states (total, site-projected, and \emph{l}-decomposed) revealed that the lower conduction bands are contributed
mainly by Hf 5\emph{d}, Ba 5\emph{d}, and 6\emph{s} states, while in the valence region, the Hf 5\emph{d} states are partially hybridized
with O 2\emph{p} and located below the near-Fermi bands
formed predominantly by O 2\emph{p} states. The bonding charge
density calculations give further evidence that the covalent bonds exist between the Hf and O atoms and the ionic bonds exist between the Ba atoms and HfO$_3$ ionic groups in BaHfO$_3$.

\section*{\ack}
This work was supported by the National Nature Science Foundation of China (No. O311011301) and the Knowledge Innovation Program of the Chinese Academy of Sciences (No. 072C201301). The authors thank the Shanghai Supercomputer Center and the Center for High Performance Computing, Northwestern Polytechnical University, from where the computational work in this paper was done and the technical support was given.



\end{document}